\documentclass[conference,compsoc]{IEEEtran}
\usepackage{graphicx} %

\title{Challenges and Future Directions in Agentic Reverse Engineering Systems }

\begin{document}
\author{
  \IEEEauthorblockN{Salem Radey, Jack West, Kassem Fawaz}
  \IEEEauthorblockA{
    \textit{University of Wisconsin-Madison} \\
    Madison, USA \\
    \{salem.radey, jwwest, kfawaz\}@wisc.edu
  }
}

\maketitle

\begin{abstract}
Agentic systems built on large language models (LLMs) are increasingly being used for complex security tasks, including binary reverse engineering (RE). Despite recent growth in popularity and capability, these systems continue to face limitations in realistic settings. Cutting-edge systems still fail in complex RE scenarios that involve obfuscation, timing, and unique architecture. In this work, we examine how agentic systems perform reverse engineering tasks with static, dynamic, and hybrid agents. Through an analysis of existing agentic tool usage, we identify several limitations, including token constraints, struggles with obfuscation, and a lack of program guardrails. From these findings, we outline current challenges and position future directions for system designers to overcome from a security perspective. 
\end{abstract}

\section{Introduction}
Agentic reverse engineering (RE) systems are collections of agents aimed to solve a given security task~\cite{shahriar2025survey}.
Each RE agent performs selected tasks in response to the system's needs at a given time.
These tasks range from red-teaming~\cite{muzsai2024hacksynthllmagentevaluation}, blue teaming~\cite{he_sentinelagent_2025}, anomaly detection~\cite{chen2025clearagent}, and many other security tasks~\cite{shahriar2025survey}.
One of the most necessary capabilities for almost all RE problems is \textit{binary RE}.

Binary RE is the process of defining a binary's functionality using its compiled code.
There are three methods in security research for which a binary can be fully explored: (1) static analysis, (2) dynamic analysis, and (3) hybrid analysis.
Static analysis attempts to define the inner workings of a binary without executing the program itself~\cite{llvmClangStatic}.
Dynamic analysis, on the other hand, is the process of analyzing the binary while it is running.
Finally, hybrid analysis combines static and dynamic analysis to analyze a binary.
Modern security research~\cite{west2024picture,cao2025jnfuzz} that uses RE often employs hybrid analysis, as it captures almost all functionality of a given binary.
Agents in a system are compartmentalized to perform one or more of the three binary types~\cite{deng_pentestgpt_nodate,ghosh_cve-llm_2025,muzsai2024hacksynthllmagentevaluation}.
However, the technical details of how these systems implement these processes are opaque.

There is no all-encompassing algorithm to RE an arbitrary binary~\cite{engel2024decidability}.
Depending on how the binary is constructed, it could completely change the tool set and approach an agent must make.
For example, Android apps require two different decompilers to fully examine a given app across two different runtimes~\cite{qian2025lamdcontextdrivenandroidmalware}.
Therefore, understanding the technical capabilities of an agentic system will define what binaries it can or cannot RE.

In this work, we aim to define the current technical details for how agentic systems perform binary RE and lay out future directions and existing challenges that must be overcome.
To define the details, we analyze open-source libraries for agentic RE systems to determine precisely how these systems perform binary RE.
We divide binary RE into three categories: static, dynamic, and hybrid analysis.
We also discuss \textbf{6} challenges and limitations that agents currently experience.

We find that agents that perform static analysis are limited in capability.
There are two main challenges agents currently face: (1) obfuscation and (2) tokenization.
Obfuscated binaries cause confusion and unneeded complexity. 
Several studies have shown that obfuscation affects agent analysis~\cite{Tan_2024} or that agents avoid obfuscated binaries entirely~\cite{dramko2025quantifying}.
However, in realistic security scenarios, obfuscated binaries are common~\cite{Tan_2024}.
Binaries, especially after decompilation, are token-heavy~\cite{udeshi_binary_2025,qian2025lamdcontextdrivenandroidmalware}.
Limited tokens constrain the coverage an agent may have.

Dynamic analysis, on the other hand, exhibits its own set of issues.
We identify three core challenges: (1) lack of dynamic analysis guardrails, (2) timeouts, and (3) reliance on emulation.
Upon reviewing codebases of existing agentic systems~\cite{enigmaPlusDebugsh,muzsai2024hacksynthllmagentevaluation,yang2024sweagentagentcomputerinterfacesenable}, we found that they commonly allow models to execute any command the agent deems necessary.
RE tools are extremely powerful pieces of software, often requiring root-level access into compartmentalized memory spaces~\cite{GitHubRadare2,frida,dynamorio}.
Recent work has shown~\cite{triedman2025multi,pasquini_hacking_2024} that it is possible to trick an agent into executing malicious code.
Thus, we argue that it may be possible to craft a binary that tricks a model into executing malicious code.

Adversarial programs commonly employ timeouts to prevent processes from being monitored~\cite{afianian2019malware}.
Given the time and latency agents naturally have, we anticipate that binaries will employ strict timeouts to block any dynamic analysis attempts.
Adversarial applications also load dynamic binaries that may be encrypted or downloaded from an off-device source, further complicating the timing of an agent's analysis.

When performing safe dynamic analysis, experts commonly load malicious programs into virtual machines (VMs).
Agents also analyze binaries within virtual environments as well for safe analysis~\cite{enigmaPlusDebugsh,yang2024sweagentagentcomputerinterfacesenable}.
However, unlike their human RE counterparts, agents can be tricked into executing malicious code~\cite{triedman2025multiagentsystemsexecutearbitrary}.
Furthermore, several virtual environments for edge devices have virtualization issues~\cite{zhou2022your}.

Finally, hybrid analysis is when an RE performs both static and dynamic analysis.
Hybrid analysis shares the core challenges of static and dynamic analyses but also exhibits unique issues.
The main challenge we identify in our research is that hybrid agents rely on a human-in-the-loop or perform all reasoning internally~\cite {shen2025pentestagent,muzsai2024hacksynthllmagentevaluation,yang2024sweagentagentcomputerinterfacesenable,enigmaPlusDebugsh}.
Both approaches have their own issues.
For example, human-in-the-loop is slow and requires an active RE to execute the instructions.
Whereas, completely automated systems reason independently and may be susceptible to adversarially crafted binaries.
We focus on automation issues since the end goal of an agentic system is to remove human dependency.

In this work, we propose several research directions for static analysis agents, informed by our expert understanding of reverse engineering.
First, we propose tokenizing static analysis based on byte patterns to lower the number of tokens the model may see.
We also posit that exploring alternative decompilation methods within an agentic system could optimize an agent's understanding of a given binary.
Both byte-pattern filtering and ML model binary translation may result in lower tokenization.
Furthermore, we argue that obfuscation must be accounted for in agentic RE systems.
Work has shown that obfuscation reduces an agent's capabilities; therefore, we must incorporate a deobfuscation component into the pipeline.

Regarding dynamic analysis, we also provide researchers with directions for future exploration to improve agent performance.
Our analysis reveals that agentic systems designers rely heavily on the model's judgment for executing commands.
This lack of guardrails could lead to system hijacking by constructing a clever binary. 
Therefore, we present two avenues for research: (1) an adversarial direction to create a binary that takes control of the host system by utilizing the lack of guardrails, and (2) building guardrails for dynamic analysis.
Another set of problems in dynamic analysis includes timeouts and the lack of sequential processing.
Both issues may limit an agent's ability to perform dynamic analysis.
Thus, researchers should design specialized tools to overcome the limitations of sequential real-time data processing for agents.

Finally, for hybrid agents, we propose a stronger reasoning methodology to prevent the execution of malicious code.
Current systems lack a voting infrastructure and safety checks to prevent an agent from performing dynamic analysis.
Hybrid systems are extremely vulnerable to malicious code as they interact with it.
Thus, further verification methods across static and dynamic analysis should be explored.

\section{Taxonomy}

In this section, we detail our taxonomy when discussing reverse engineering.
Our focus will be on agents that can analyze a binary on a given device.
Meaning, the agent has full access to a binary and unlimited resources to perform any type of analysis.

\subsection{Reverse Engineering}
We examine papers that attempt to automate \textit{reverse engineering} using agents.
We define reverse engineering as the process of iteratively interacting with an arbitrary computer program.
Specifically, we focus on reverse-engineering a given binary across multiple architectures.
Therefore, our analysis will exclude reverse engineering that attempts to exploit binaries off a device on which the given agent exists.

Reverse engineering can be described into three different types: (1) static, (2) dynamic, and (3) hybrid analysis. 
Static analysis is the process of examining a binary without executing it on the device.
Dynamic analysis, on the other hand, is analyzing a binary while it is running.
Hybrid analysis combines static and dynamic analysis to evaluate a binary.

\subsection{AI Agent and Agentic Scope}

We define an AI agent as an LLM designed to solve a task, equipped with the tools to do so~\cite{shen2025pentestagent}.
LLM agents plan and execute their own individual decisions by observing environmental feedback~\cite{shahriar2025survey}.
Agentic systems are collections of agents aiming to work together to accomplish a goal.
In this work, we examine an agentic system where the goal is to reverse engineer a given binary.

Rather than defining a threat model, we define the capabilities and goals of an arbitrary agent.
First, we assume that the agent has no access to the source code of a given binary.
We argue that the source code of a binary greatly simplifies RE tasks and is unrealistic for real-world security analysis.
Our focus will purely be on local binary analysis.
Some RE tasks require networking capabilities, such as port scanning or network traffic analysis.
We do not discuss any RE task that requires networking.
We assume no restrictions on the size, target device, and capabilities for a given binary.

In all agents and agentic systems, we observed that they all decompile binaries.
Decompiled binaries attempt to translate the bytes into a more human-readable form.
Tools such as Ghidra~\cite{githubGhidra} and IDA~\cite{IDAproWebsite} translate binary code into a higher-level language.
Agents that study decompiled binaries first preprocess the binaries to make them easier to read.
We do not restrict agentic systems to a set of tools; any decompilation method is valid.

\section{Overview of Agents}

This section presents a technical analysis of how agents perform various types of binary reverse engineering.
Our focus is on the \textit{technical capabilities} of the agents rather than their overall performance or goal.
Several surveys have explored the diverse applications of LLM agents~\cite{shahriar2025survey,yaacoub_large_2025,ren_large_nodate,hu_sok_2025}, but our work focuses on their use in binary analysis from a security perspective.
The focus on technical capabilities is intended to clarify how these agents perform binary reverse engineering.

\begin{figure}[t]
    \centering
    \includegraphics[width=\columnwidth]{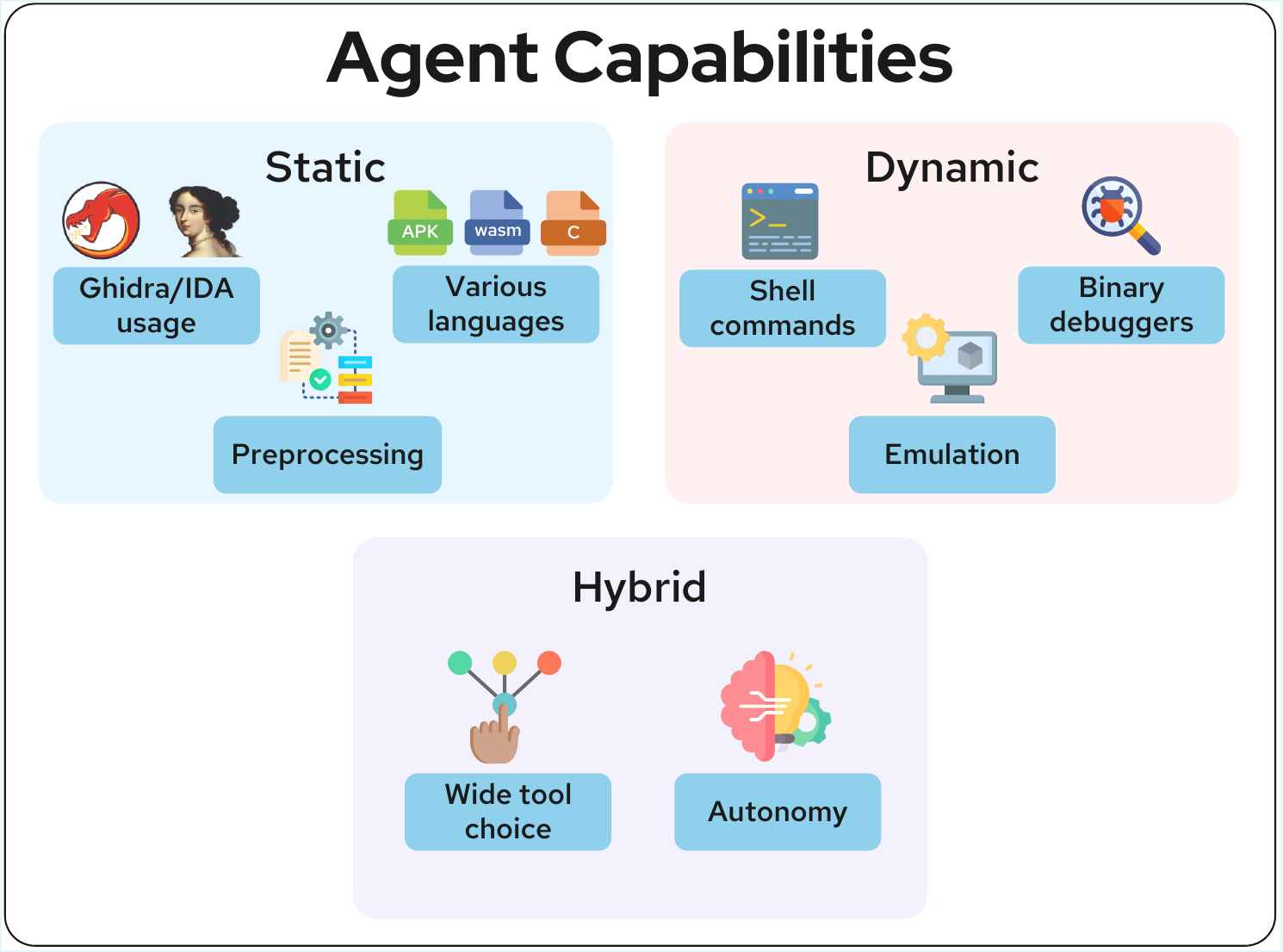}
    \caption{An overview of agent capabilities by type.}
    \label{fig:agents_fig}
\end{figure}

We segment agents into two categories: static analysis and hybrid analysis.
Static analysis agents are common because static analysis can assist with a wide range of tasks beyond RE.
Therefore, the community's understanding of static analysis agents exceeds that of hybrid agents. 
Our literature review found that no agent focuses on dynamic analysis; they would perform hybrid analysis.
Thus, our discussion pertaining to hybrid analysis will detail current dynamic analysis capabilities. An overview of these capabilities can be seen in Figure~\ref{fig:agents_fig}.

\subsection{Static Analysis Agents}

Static analysis is an essential component of any binary reverse engineering pipeline and is, therefore, commonly performed by agents~\cite{deng_pentestgpt_nodate,shen2025pentestagent,pasquini_hacking_2024,hu_compileagent_nodate,chen_recopilot_2025,ghosh_cve-llm_2025,kong_vulnbot_2025,lin_ircopilot_2025,zhuo_training_2025,balassone_cybersecurity_2025,udeshi_binary_2025}.
We found that agents perform static analysis on decompiled binaries.

\subsubsection{Decompiled Code}
Binary decompilation is a common practice in reverse engineering~\cite{wang2025salt4decompileinferringsourcelevelabstract}.
The purpose of decompilation is to provide the human reverse engineer a readable interpretation of a given binary.
The human reverse engineer then reads and interprets the outputs to determine the program's goal.
We find that decompilation is most commonly performed using Ghidra~\cite{githubGhidra} or IDA Pro~\cite{IDAproWebsite}. However, some studies also seek to train LLM agents to perform decompilation. Tan et al.~\cite{Tan_2024} introduce LLM4Decompile, the first series of open source LLM trained for decompilation. They introduce both LLM4Decompile-End models which do decompilation directly, as well as LLM4Decompile-Ref models which refine Ghidra output.  LLM4Decompile-End uses \texttt{ObjDump} to disassemble binaries prior to decompilation. 

Likewise, Rong et al.~\cite{rong2024disassemblingobfuscatedexecutablesllm} present \textsc{disasLLM}, an end-to-end LLM disassembler for highly obfuscated executables. The system performs an initial disassembly and then uses an LLM-based classifier to validate decoded instructions. To improve efficiency, \textsc{disasLLM} batches LLM queries and outputs the corrected disassembly after all instructions are checked. Unlike LLM4Decompile, \textsc{disasLLM} is end-to-end and does not interact with traditional decompilers such as Ghidra or IDA Pro. 

Other works explore web runtimes and languages such as WebAssembly (WASM). She et al.~\cite{she2024wadecDecompilingWebassembly} present \textsc{WaDec}, an approach that converts Wasm binaries to the textual \texttt{wat} format and performs block-wise decompilation using a fine-tuned LLM. By segmenting functions into Wasm loop blocks and applying a unified variable renaming scheme, WaDec improves scalability and readability, producing decompiled code with minimal code inflation. 

Other agents turn to JADx~\cite{GitHubSkylotjadx} when reverse engineering Android dex and APK files. Qian et al.~\cite{qian2025lamdcontextdrivenandroidmalware} introduce LAMD: an LLM-powered framework that utilizes key context extraction and tier-wise code reasoning to detect malware more accurately than standalone LLMs. The LAMD framework uses existing models, with the paper evaluating GPT-4o-mini. However, since GPT-4o-mini and many other LLM cannot process APKs directly, the authors use JADx to decompile the code as a pre-processing step before sending it to the LLM. Since Android APK files consist of both C and Java code~\cite{androidApplicationFundamentals}, there is additional pre-processing overhead to account for the differing methods required to decompile each language.  

Most studies tend to use IDA Pro or Ghidra. Shang et al.~\cite{shang2025binmetriccomprehensivebinaryanalysis} introduce BinMetric, a benchmark of 1000 questions pertaining to common real-world binary analysis tasks. To obtain data to feed the LLM, the authors compile and strip the source code, then use IDA Pro to decompile and disassemble the compiled binaries. Although they choose to use IDA Pro, the authors note the prevalence of tools such as Ghidra, Radare2~\cite{GitHubRadare2}, and Clang Static Analyzer~\cite{llvmClangStatic} as static analysis tools. However, they note that the tools require significant manual effort and expert knowledge to use. Additionally, the authors use the SrcML tool~\cite{Maletic_Collard_2015} to extract information from the source files and convert it into XML format as additional data.

\subsection{Hybrid Analysis}

Hybrid analysis combines static and dynamic analysis to leverage the strengths of both approaches when reverse engineering. We define a hybrid analysis agent as an agent that can use tools for both static and dynamic analysis.
All agents we observed that perform dynamic analysis do so in tandem with static analysis.
In this section, we provide an overview of dynamic agent capabilities and discuss hybrid implementations.

\subsubsection{Dynamic Analysis Agents}

Dynamic analysis is the process of observing code behavior at runtime. Due to dynamic code loading~\cite{afianian2019malware}, many behaviors of a program are not inferable through static analysis and are only observed during runtime. To observe these behaviors, reverse engineers use tools for dynamic analysis, including debuggers and dynamic instrumentation. Dynamic analysis allows a reverse engineer to define complex behaviors using captured values, execution traces, and breakpoints.

Despite the advantages of dynamic analysis, it remains more difficult to scale than static analysis. Currently, the interactive and adaptive components of dynamic analysis make it difficult for LLMs to perform independently. Therefore, human analysts must still account for a large portion of the work~\cite{basquedecompiling}. Additionally, pipelines used for dynamic analysis are tightly coupled to the code's architecture, making portability a challenge~\cite{west2024picture,afianian2019malware,cao2025jnfuzz}. 

An example tool used for dynamic analysis is the GNU Debugger (GDB)~\cite{sourcewareGDBProject,zhuo2025cyberzerotrainingcybersecurityagents,abramovich2025enigmainteractivetoolssubstantially}. 
GDB allows for step-by-step instruction-level analysis of a binary. 
Beyond stepping through code, GDB also features disassembly, which can be done at runtime, decompilation, the ability to view the contents of registers, and breakpoints~\cite{ctf101GDB}. 
Zhuo et al. ~\cite{zhuo2025traininglanguagemodelagents} present CTF-Dojo, an executable runtime designed to train cybersecurity agents using CTF challenges. 
CTF-Dojo builds upon EnIGMA+~\cite{githubEnIGMAplus, zhuo2025cyberzerotrainingcybersecurityagents} for its agent, which implements dynamic reverse engineering capabilities using GDB. 
EnIGMA+ implements GDB only with the commands \texttt{break}, which adds a breakpoint, \texttt{continue}, which continues execution of the binary, \texttt{stepi}, which steps to a specific line, and \texttt{debug\_exec}, which allows for arbitrary GDB command execution\cite{enigmaPlusDebugsh}. Disassembly and decompilation are handled separately with Ghidra. 

Likewise, Muzsai et al.~\cite{muzsai2024hacksynthllmagentevaluation} introduce HackSynth, an LLM agent capable of autonomous penetration testing. HackSynth’s architecture consists of a planner and summarizer that iteratively generate and interpret bash commands within a containerized Linux environment. Unlike CTF-Dojo, HackSynth uses \textit{any} bash command available to the agent. 

\subsubsection{Hybrid Implementations}
While dynamic analysis agents appear only in hybrid systems, there are many ways to design a hybrid agent.
There are two methods to perform hybrid analysis: (1) human-in-the-loop and (2) fully autonomous.
Human-in-the-loop systems execute the agents' instructions and verify their feasibility and security.
The humans in this process are expected to run the commands so that the agent can execute commands safely.
Whereas autonomous systems are left completely to their own devices when performing analysis.
Meaning, all of the reasoning for RE comes from the collection of agents communicating.

Deng et al.~\cite{deng_pentestgpt_nodate} present \textsc{PentestGPT}, an LLM-driven framework for automated pentesting. Similar to HackSynth, \textsc{PentestGPT} comprises multiple agents. Whereas HackSynth used a planner and summarizer, \textsc{PentestGPT} employs reasoning, generation, and parsing agents.
\textsc{PentestGPT} does not enable tool usage; rather, it operates with a human-in-the-loop strategy, in which a human expert serves solely as an executor, strictly following the LLM’s suggestions.
\textsc{PentestGPT} represents a human-in-the-loop system whereas HackSynth is completely automated.

Another case of an agentic hybrid approach is by Pasquini et al.~\cite{pasquini_hacking_2024}, which creates MANTIS: a defense against LLM-driven cyberattacks. MANTIS operates autonomously, responds to detected anomalous interactions, and employs both active (dynamic) and passive (static) defenses. The static component analyzes the attacker’s goals, constraints, and control logic to misdirect them. The dynamic component instruments a counterattack by leading the adversarial LLM to open a reverse shell on the adversary's machine.

\section{Current Challenges}
Reverse engineers face a broad range of challenges that stem from the increasing complexity of modern software. Obfuscation techniques, large-scale code bases, and dynamic runtime behavior all serve to complicate RE. These methods fundamentally limit the understanding an analyst can glean from a program's bytes, structure, and behavior. As with many technical disciplines, reverse engineering is continually evolving. As RE improves, so do anti-RE techniques hindering models. 

\subsection{Static Analysis Challenges}
Static reverse engineering comes with a set of inherent challenges that analysts face regardless of their expertise. Techniques such as code obfuscation through symbol stripping, variable name changes, control flow changes, and comment removal obscure code semantics and impede understanding for both human and LLM reverse engineers. Additionally, the code bases of reversing targets are often large, which requires additional time and understanding to rationalize. 

\subsubsection{Obfuscation}
Just as in non-agentic reverse engineering, obfuscation remains a hurdle for agentic systems.
For agentic systems, obfuscation provides new challenges. Since LLM-based agents rely on their ability to detect syntax patterns they have been trained to recognize, previously unseen or unconventional obfuscation techniques disrupt the agents' ability to function optimally~\cite{Tan_2024}. This increases the likelihood of misclassifications or an incorrect prioritization of code importance. 

Tan et al.~\cite{Tan_2024} find that Control Flow Flattening (CFF) and Bogus Control Flow (BCF) both decrease the success rate of decompilation by over 70\%, showing that even with common obfuscation techniques, there exists a drastic decrease in reversing accuracy. This implies that complex obfuscation techniques could drop accuracies \textit{even lower}.

As a result, many works dealing with binary code choose to forgo including obfuscated binaries altogether, since they pose challenges for LLMs~\cite{shang2024fargonebinarycode, shang2025binmetriccomprehensivebinaryanalysis, basquedecompiling}. However, closed-source real-world applications commonly employ obfuscation techniques, meaning that when given an obfuscated binary, these agents are likely to fail. 

\subsubsection{Tokenization}
LLM-based agents operate with a bounded context depending on their token limit, which constrains the amount of decompiled code that can be considered at once~\cite{qian2025lamdcontextdrivenandroidmalware}. The natural workaround for this is to segment the input into chunks that fit the token limit, then analyze them sequentially. However, this implementation fails to account for long control flows that span the entire code and for functions or variables defined outside the current section. This lack of context impairs the agent's ability to properly analyze the code's relationships as a whole. 

Udeshi et al.~\cite{udeshi_binary_2025} propose a framework using LLMs to generate summarizations of binary diffs for malware detection in supply chain scenarios. The study was conducted on 104 binaries using five LLMs and required token usage ranging from 100M to 500M per model. This study represents a best-case scenario, where the authors removed exact matches from the diff, thereby reducing token counts. However, this removal may affect an agent's ability to fully understand the context of a binary. Current work demonstrates that larger binaries require a balance between tokens and functionality. 

\subsection{Dynamic Analysis Challenges}
Dynamic analysis poses challenges for agentic reverse-engineering systems from both architectural and tooling perspectives. Current works demonstrate an over-reliance on shell commands, which may constrain the agent's ability to perform more diverse runtime analysis and, depending on the implementation, introduce security risks. Additionally, LLM-based agents are hindered by their prolonged reasoning times, which may impact time-sensitive interactions and lead reverse engineers to reach incorrect conclusions due to timeouts and other unintended behaviors. In addition, low-level emulation environments are hard to emulate, limiting the dynamic analysis an agent can perform on niche hardware.

\subsubsection{Lack of Guardrails}
Some agents lack guardrails for the commands they are allowed to execute. One example of this is CTF-Dojo by Zhuo et al.~\cite{zhuo2025traininglanguagemodelagents} which builds upon \textsc{Enigma+}~\cite{githubEnIGMAplus}. \textsc{Enigma+} allows unrestricted dynamic analysis on a given binary. This means an attacker could purposefully trick the agent into executing unsafe commands on the device. Another example is Muzsai et al.~\cite{muzsai2024hacksynthllmagentevaluation}, who developed HackSynth, an LLM-based pentesting agent tested on CTF challenges. Similar to CTF-Dojo, HackSynth allows commands to be input without validation.

\subsubsection{Timeouts}
A key challenge for executing dynamic analysis with agentic systems arises from the variable processing times of LLMs. LLMs utilize multi-step reasoning and planning, which may require an extended period of time to 'think'. This inconsistency in execution time can trigger timeouts~\cite{afianian2019malware} or miss key interactions altogether. Reverse engineers interacting with these systems may draw spurious conclusions from these timeouts, assuming the agent is operating under proper conditions. 

\subsubsection{Reliance on Emulation}
Almost all existing agentic systems perform analysis on desktops.
These systems use Docker and/or QEMU to create virtual machines for arbitrary devices and architectures.
However, several edge devices are notoriously hard to simulate~\cite{zhou2022your}.
Furthermore, some operating systems require multiple application components compiled with different compilers~\cite{qian2025lamdcontextdrivenandroidmalware}.
To our knowledge, edge device RE is not currently being explored due to emulation restrictions.

\subsection{Hybrid Analysis Challenges}
Hybrid analysis shares challenges with static and dynamic analysis.
However, hybrid systems exhibit one major challenge regardless of technical capabilities: system reasoning.
Binaries are not definable deterministically~\cite{engel2024decidability}.
Therefore, unchecked reasoning in agentic systems is more likely to lead to failure on RE tasks for complex applications.
For example, if an agent hyperfocuses on a specific region of the binary, which may be obvious to a human, it could waste hours of compute time without any human intervention.

Another issue found in hybrid systems is the reliance on a human-in-the-loop.
Human-in-the-loop adds time to the RE process and limits the model's capabilities while retaining security and guidance.
Interestingly, the challenges of full automation and Human-in-the-loop mirror each other.
In this section, we describe the challenges associated with automation, as the goal for the community is automation rather than a human-in-the-loop approach.

\subsubsection{Never Ending Automation}
As with any cat-and-mouse chase in security, the attacker adapts.
If agentic systems attempt to fully automate RE practices, we foresee adversarial binary development to maximize the agent's processing time.
For example, new obfuscation techniques could emerge that force agents to triage long execution chains to extract a single piece of information.
Rather than optimizing for human deobfuscation, the adversary will likely obfuscate to maximize hindrance to agentic communication and time.

\section{Future Directions}

Agentic RE is still young.
Many systems and agents are still vulnerable.
Secure systems, including agentic ones, require solutions for edge cases to prevent system compromise.
Below, we outline several research directions to address existing challenges.

\subsection{Static Analysis Directions}
Our analysis indicates that static analysis agents are the most common.
Agentic systems have several methods for understanding and interacting with decompiled code from a given binary.
However, several challenges exist.
In this section, we detail potential ways to improve static analysis agents from a security researcher's perspective.

\subsubsection{Raw Binary Analysis}
Currently, no agents deal with raw binary inputs.
This is by design, as agents work with natural languages rather than bits and bytes.
However, adversaries can exploit reliance on decompilation tools to hinder agentic reverse engineering.
Since Ghidra and IDA deploy their own decompilation algorithms, it is possible to exploit the decompilation process to hide functionality~\cite{qin2021idev}.
Qin et al.~\cite{qin2021idev} demonstrate an attack where an adversary can design assembly instructions that create a semantic deviation.
One of those types is \textit{disassembly deviation}, where, depending on the compiler, a set of instructions may yield different results.
Thus, future work should explore alternative methods for binary decompilation such that they avoid or detect decompilation evasion.

\subsubsection{Add Deobfuscation to Agentic Pipelines}
Obfuscation is a challenge in RE for both scalable~\cite{cao2025jnfuzz} and manual methods~\cite{west2024picture}.
Agentic frameworks are no different~\cite{chen2025clearagent,chen_recopilot_2025}, with many frameworks forgoing~\cite{chen_recopilot_2025} or lightly deobfuscating~\cite{chen2025clearagent}.
Thus, we argue that research should be done to allow agents to perform deobfuscation.
Deobfuscating a binary will likely increase the accuracy of malware detection for agentic systems, as the system will spend less time understanding instructions.

\subsubsection{Clever Tokenization}
We found that most agentic systems that perform decompilation using Ghidra/IDA feed the output directly to the agent~\cite{deng_pentestgpt_nodate}.
The problem with feeding decompilations to agents is that they contain repetitive information (e.g., stack frame construction, local variable allocation) that is irrelevant to the high-level functionality.
Humans actively ignore repetitive functionality and focus on core logic for understanding the binary.
LLM agents, on the other hand, may not ignore irrelevant blocks of instructions, which can lead to large tokenized decompilations and limit their capabilities.
Some works, such as Udeshi et al.~\cite{udeshi_binary_2025}, cleverly leverage the fact that the agent is likely aware of open-source projects and can remove irrelevant functionality.
While a clever solution, if presented with a binary that the system is unaware of, it may fail.

\subsection{Dynamic Analysis Directions}
Work has shown that manipulating agents performing dynamic analysis is possible~\cite{pasquini_hacking_2024}.
We anticipate that these attacks will become more complex as researchers explore techniques to interfere with agentic RE.
This section describes potential solutions to the challenges in dynamic analysis. 

\subsubsection{Lack of Guardrails}
One common theme among the agents we review is a lack of guardrails on their capabilities when exploring a binary.
Binary analysis tools are extremely powerful and may allow for a cleverly crafted binary to take control of a system.
We propose that researchers attack agents during the dynamic analysis phase, as the binary serves as a dynamic adversary and can adjust its behavior in real time.
It is expected that an adversarial binary has situational awareness and can adapt to human RE efforts.

\subsubsection{Timeouts}
Adversarial binaries employ strict timeouts when executing critical tasks to prevent dynamic analysis~\cite{afianian2019malware}.
Agents are more susceptible to timeouts than their human counterparts due to network latency and limited thinking time.
A binary built to attack high-resource systems could wrap important functionality around strict thread timeouts that easily defeat thinking models.
A solution to this problem is for the agent to control the time-estimation mechanisms in the observed binary.
However, this solution requires precise function hooks and control over which time value the program observes.
Timeouts offer a unique opportunity to develop \textit{agent-specific RE tools} that build around natural weaknesses agents have.
If a human RE built a generalized tool to control the clock based on an agent's response time, the agent could automatically adjust the clock to account for its thinking time while the binary remains unaware.

\subsubsection{Improving Emulation Environments}
While the technology does not currently exist, as a community, we should build stronger emulation environments for agentic systems.
Several components could hypothetically be simulated using Gem5~\cite{gem5gem5Gem5}, enabling deeper security analysis of hardware systems.
Therefore, we propose building a similar tool that evaluates binaries on lesser-known hardware components that are challenging to emulate.
This hypothetical tool will allow for scalable IoT analysis by an agentic system.

\subsection{Hybrid Analysis Directions}
We argue that agents designed for hybrid analysis are susceptible to all of the challenges above as well as being vulnerable to their own.

\subsubsection{Protection from Infinite Tool Calls}
We need to design protections from novel obfuscation schemes designed to impede hybrid agents.
The goal of the hypothetical scheme would be to force an agent to waste its time potentially indefinitely.
While this does not yet exist, we argue that it may in the near future and must be addressed to ensure robustness and convergence.

\section{Conclusion}

In this work, we examine how agentic systems perform binary reverse engineering across static, dynamic, and hybrid settings. Additionally, we identify \textbf{6} challenges within the space. For static analysis, we identify challenges with obfuscation and tokenization. To address these challenges, we propose a future research attempt to perform raw binary analysis with agents and to add deobfuscation to agentic pipelines. For dynamic analysis, we note a lack of guardrails, issues stemming from timeouts, and a reliance on emulation. We outline future directions for binary design defenses, clock-based timeout protection, and improved emulation environments. Finally, we describe how hybrid agents may lead to an infinite sequence of tool calls. Our solution is to design defenses against infinite-agentic RE such that future obfuscation techniques will always converge.

\bibliographystyle{IEEEtran}
\bibliography{references}

@misc{zhuo2025traininglanguagemodelagents,
      title={Training Language Model Agents to Find Vulnerabilities with CTF-Dojo}, 
      author={Terry Yue Zhuo and Dingmin Wang and Hantian Ding and Varun Kumar and Zijian Wang},
      year={2025},
      eprint={2508.18370},
      archivePrefix={arXiv},
      primaryClass={cs.SE},
      url={https://arxiv.org/abs/2508.18370}, 
}

@misc{githubEnIGMAplus,
	author = {},
	title = {{G}it{H}ub - amazon-science/{C}yber-{Z}ero: {C}yber-{Z}ero: {T}raining {C}ybersecurity {A}gents {W}ithout {R}untime --- github.com},
	howpublished = {\url{https://github.com/amazon-science/Cyber-Zero}},
	year = {2025}
}

@misc{zhuo2025cyberzerotrainingcybersecurityagents,
      title={Cyber-Zero: Training Cybersecurity Agents without Runtime}, 
      author={Terry Yue Zhuo and Dingmin Wang and Hantian Ding and Varun Kumar and Zijian Wang},
      year={2025},
      eprint={2508.00910},
      archivePrefix={arXiv},
      primaryClass={cs.CR},
      url={https://arxiv.org/abs/2508.00910}, 
}

@misc{githubGhidra,
	author = {},
	title = {{G}it{H}ub - {N}ational{S}ecurity{A}gency/ghidra: {G}hidra is a software reverse engineering ({S}{R}{E}) framework --- github.com},
	howpublished = {\url{https://github.com/NationalSecurityAgency/ghidra}},
	year = {}
}

@inproceedings{shen2025pentestagent,
  title={Pentestagent: Incorporating llm agents to automated penetration testing},
  author={Shen, Xiangmin and Wang, Lingzhi and Li, Zhenyuan and Chen, Yan and Zhao, Wencheng and Sun, Dawei and Wang, Jiashui and Ruan, Wei},
  booktitle={Proceedings of the 20th ACM Asia Conference on Computer and Communications Security},
  pages={375--391},
  year={2025}
}

@article{shahriar2025survey,
  title={A Survey on Agentic Security: Applications, Threats and Defenses},
  author={Shahriar, Asif and Rahman, Md Nafiu and Ahmed, Sadif and Sadeque, Farig and Parvez, Md Rizwan},
  journal={arXiv preprint arXiv:2510.06445},
  year={2025}
}

@article{yaacoub_large_2025,
	title = {Large language models: applications, limitations, challenges, and recommendations in cybersecurity, digital forensics, and ethical hacking},
	issn = {1958-9395},
	shorttitle = {Large language models},
	url = {https://doi.org/10.1007/s12243-025-01134-9},
	doi = {10.1007/s12243-025-01134-9},
	language = {en},
	urldate = {2026-01-05},
	journal = {Annals of Telecommunications},
	author = {Yaacoub, Jean Paul A. and Noura, Hassan N. and Salman, Ola and Pujolle, Guy},
	month = nov,
	year = {2025},
}

@misc{pasquini_hacking_2024,
	title = {Hacking {Back} the {AI}-{Hacker}: {Prompt} {Injection} as a {Defense} {Against} {LLM}-driven {Cyberattacks}},
	shorttitle = {Hacking {Back} the {AI}-{Hacker}},
	url = {http://arxiv.org/abs/2410.20911},
	doi = {10.48550/arXiv.2410.20911},
	language = {en},
	urldate = {2026-01-05},
	publisher = {arXiv},
	author = {Pasquini, Dario and Kornaropoulos, Evgenios M. and Ateniese, Giuseppe},
	month = nov,
	year = {2024},
	note = {arXiv:2410.20911 [cs]},
}

@article{hu_compileagent_nodate,
	title = {{CompileAgent}: {Automated} {Real}-{World} {Repo}-{Level} {Compilation} with {Tool}-{Integrated} {LLM}-based {Agent} {System}},
	language = {en},
	author = {Hu, Li and Chen, Guoqiang and Shang, Xiuwei and Cheng, Shaoyin and Wu, Benlong and Li, Gangyang and Zhu, Xu and Zhang, Weiming and Yu, Nenghai},
}

@article{ghosh_cve-llm_2025,
	title = {{CVE}-{LLM}: {Ontology}-{Assisted} {Automatic} {Vulnerability} {Evaluation} {Using} {Large} {Language} {Models}},
	volume = {39},
	copyright = {Copyright (c) 2025 Association for the Advancement of Artificial Intelligence},
	issn = {2374-3468},
	shorttitle = {{CVE}-{LLM}},
	url = {https://ojs.aaai.org/index.php/AAAI/article/view/35139},
	doi = {10.1609/aaai.v39i28.35139},
	language = {en},
	number = {28},
	urldate = {2026-01-05},
	journal = {Proceedings of the AAAI Conference on Artificial Intelligence},
	author = {Ghosh, Rikhiya and Stockhausen, Hans-Martin von and Schmitt, Martin and Vasile, George Marica and Karn, Sanjeev Kumar and Farri, Oladimeji},
	month = apr,
	year = {2025},
	pages = {28757--28765},
}

@misc{kong_vulnbot_2025,
	title = {{VulnBot}: {Autonomous} {Penetration} {Testing} for {A} {Multi}-{Agent} {Collaborative} {Framework}},
	shorttitle = {{VulnBot}},
	url = {http://arxiv.org/abs/2501.13411},
	doi = {10.48550/arXiv.2501.13411},
	language = {en},
	urldate = {2026-01-05},
	publisher = {arXiv},
	author = {Kong, He and Hu, Die and Ge, Jingguo and Li, Liangxiong and Li, Tong and Wu, Bingzhen},
	month = jan,
	year = {2025},
	note = {arXiv:2501.13411 [cs]},
}

@article{ren_large_nodate,
	title = {Large {Language} {Models} for {Cybersecurity} {Intelligence}, {Threat} {Hunting}, and {Decision} {Support}},
	language = {en},
	author = {Ren, Shaochen and Chen, Shiyang},
}

@article{deng_pentestgpt_nodate,
	title = {{PentestGPt}: {Evaluating} and {Harnessing} {Large} {Language} {Models} for {Automated} {Penetration} {Testing}},
	language = {en},
	author = {Deng, Gelei and Liu, Yi and Robotics, Alias and Klagenfurt, Alpen-Adria-Universität and Liu, Peng and Li, Yuekang and Zhang, Tianwei and Liu, Yang and Klagenfurt, Alpen-Adria-Universität and Rass, Stefan},
}

@misc{lin_ircopilot_2025,
	title = {{IRCopilot}: {Automated} {Incident} {Response} with {Large} {Language} {Models}},
	shorttitle = {{IRCopilot}},
	url = {http://arxiv.org/abs/2505.20945},
	doi = {10.48550/arXiv.2505.20945},
	language = {en},
	urldate = {2026-01-06},
	publisher = {arXiv},
	author = {Lin, Xihuan and Zhang, Jie and Deng, Gelei and Liu, Tianzhe and Zhang, Tianwei and Guo, Qing and Chen, Riqing},
	month = oct,
	year = {2025},
	note = {arXiv:2505.20945 [cs]},
}

@misc{he_sentinelagent_2025,
	title = {{SentinelAgent}: {Graph}-based {Anomaly} {Detection} in {Multi}-{Agent} {Systems}},
	shorttitle = {{SentinelAgent}},
	url = {http://arxiv.org/abs/2505.24201},
	doi = {10.48550/arXiv.2505.24201},
	language = {en},
	urldate = {2026-01-06},
	publisher = {arXiv},
	author = {He, Xu and Wu, Di and Zhai, Yan and Sun, Kun},
	month = may,
	year = {2025},
	note = {arXiv:2505.24201 [cs]},
}

@misc{chen_recopilot_2025,
	title = {{ReCopilot}: {Reverse} {Engineering} {Copilot} in {Binary} {Analysis}},
	shorttitle = {{ReCopilot}},
	url = {http://arxiv.org/abs/2505.16366},
	doi = {10.48550/arXiv.2505.16366},
	language = {en},
	urldate = {2026-01-06},
	publisher = {arXiv},
	author = {Chen, Guoqiang and Sun, Huiqi and Liu, Daguang and Wang, Zhiqi and Wang, Qiang and Yin, Bin and Liu, Lu and Ying, Lingyun},
	month = may,
	year = {2025},
	note = {arXiv:2505.16366 [cs]},
}

@misc{balassone_cybersecurity_2025,
	title = {Cybersecurity {AI}: {Evaluating} {Agentic} {Cybersecurity} in {Attack}/{Defense} {CTFs}},
	shorttitle = {Cybersecurity {AI}},
	url = {http://arxiv.org/abs/2510.17521},
	doi = {10.48550/arXiv.2510.17521},
	language = {en},
	urldate = {2026-01-06},
	publisher = {arXiv},
	author = {Balassone, Francesco and Mayoral-Vilches, Víctor and Rass, Stefan and Pinzger, Martin and Perrone, Gaetano and Romano, Simon Pietro and Schartner, Peter},
	month = oct,
	year = {2025},
	note = {arXiv:2510.17521 [cs]},
}

@misc{hu_sok_2025,
	title = {{SoK}: {Potentials} and {Challenges} of {Large} {Language} {Models} for {Reverse} {Engineering}},
	shorttitle = {{SoK}},
	url = {http://arxiv.org/abs/2509.21821},
	doi = {10.48550/arXiv.2509.21821},
	language = {en},
	urldate = {2026-01-06},
	publisher = {arXiv},
	author = {Hu, Xinyu and Fu, Zhiwei and Xie, Shaocong and Ding, Steven H. H. and Charland, Philippe},
	month = sep,
	year = {2025},
	note = {arXiv:2509.21821 [cs]},
}

@misc{udeshi_binary_2025,
	title = {Binary {Diff} {Summarization} using {Large} {Language} {Models}},
	url = {http://arxiv.org/abs/2509.23970},
	doi = {10.48550/arXiv.2509.23970},
	language = {en},
	urldate = {2026-01-06},
	publisher = {arXiv},
	author = {Udeshi, Meet and Putrevu, Venkata Sai Charan and Krishnamurthy, Prashanth and Anantharaman, Prashant and Carrick, Sean and Karri, Ramesh and Khorrami, Farshad},
	month = sep,
	year = {2025},
	note = {arXiv:2509.23970 [cs]},
}

@misc{zhuo_training_2025,
	title = {Training {Language} {Model} {Agents} to {Find} {Vulnerabilities} with {CTF}-{Dojo}},
	url = {http://arxiv.org/abs/2508.18370},
	doi = {10.48550/arXiv.2508.18370},
	language = {en},
	urldate = {2026-01-07},
	publisher = {arXiv},
	author = {Zhuo, Terry Yue and Wang, Dingmin and Ding, Hantian and Kumar, Varun and Wang, Zijian},
	month = sep,
	year = {2025},
	note = {arXiv:2508.18370 [cs]},
}

@misc{ctf101GDB,
	author = {OSIRIS Lab \& CTFd LLC},
	title = {{D}ebuggers - {C}{T}{F} {H}andbook --- ctf101.org The GNU Debugger (GDB)},
	howpublished = {\url{https://ctf101.org/reverse-engineering/what-is-gdb/}},
	year = {2024}
}

@misc{GitHubSkylotjadx,
	author = {Skylot},
	title = {{G}it{H}ub - skylot/jadx: {D}ex to {J}ava decompiler --- github.com},
	howpublished = {\url{https://github.com/skylot/jadx}},
	year = {}
}

@misc{qian2025lamdcontextdrivenandroidmalware,
      title={LAMD: Context-driven Android Malware Detection and Classification with LLMs}, 
      author={Xingzhi Qian and Xinran Zheng and Yiling He and Shuo Yang and Lorenzo Cavallaro},
      year={2025},
      eprint={2502.13055},
      archivePrefix={arXiv},
      primaryClass={cs.CR},
      url={https://arxiv.org/abs/2502.13055}, 
}

@misc{shang2025binmetriccomprehensivebinaryanalysis,
      title={BinMetric: A Comprehensive Binary Analysis Benchmark for Large Language Models}, 
      author={Xiuwei Shang and Guoqiang Chen and Shaoyin Cheng and Benlong Wu and Li Hu and Gangyang Li and Weiming Zhang and Nenghai Yu},
      year={2025},
      eprint={2505.07360},
      archivePrefix={arXiv},
      primaryClass={cs.SE},
      url={https://arxiv.org/abs/2505.07360}, 
}

@misc{Maletic_Collard_2015, 
title={Exploration, Analysis, and Manipulation of Source Code Using srcML}, 
url={http://dx.doi.org/10.1109/ICSE.2015.302}, 
DOI={10.1109/icse.2015.302}, 
journal={2015 IEEE/ACM 37th IEEE International Conference on Software Engineering}, publisher={IEEE}, 
author={Maletic, Jonathan I. and Collard, Michael L.}, 
year={2015}, 
month=may }

@inproceedings{Tan_2024,
   title={LLM4Decompile: Decompiling Binary Code with Large Language Models},
   url={http://dx.doi.org/10.18653/v1/2024.emnlp-main.203},
   DOI={10.18653/v1/2024.emnlp-main.203},
   booktitle={Proceedings of the 2024 Conference on Empirical Methods in Natural Language Processing},
   publisher={Association for Computational Linguistics},
   author={Tan, Hanzhuo and Luo, Qi and Li, Jing and Zhang, Yuqun},
   year={2024},
   pages={3473–3487} }

@misc{shang2024fargonebinarycode,
      title={How Far Have We Gone in Binary Code Understanding Using Large Language Models}, 
      author={Xiuwei Shang and Shaoyin Cheng and Guoqiang Chen and Yanming Zhang and Li Hu and Xiao Yu and Gangyang Li and Weiming Zhang and Nenghai Yu},
      year={2024},
      eprint={2404.09836},
      archivePrefix={arXiv},
      primaryClass={cs.SE},
      url={https://arxiv.org/abs/2404.09836}, 
}

@misc{IDAproWebsite,
	author = {Hex-Rays},
	howpublished = {\url{https://hex-rays.com/ida-pro}},
	year = {}
}

@article{basquedecompiling,
  title={Decompiling the Synergy: An Empirical Study of Human--LLM Teaming in Software Reverse Engineering},
  author={Basque, Zion Leonahenahe and Doria, Samuele and Soneji, Ananta and Gibbs, Wil and Doup{\'e}, Adam and Shoshitaishvili, Yan and Losiouk, Eleonora and Wang, Ruoyu and Aonzo, Simone}
}

@misc{muzsai2024hacksynthllmagentevaluation,
      title={HackSynth: LLM Agent and Evaluation Framework for Autonomous Penetration Testing}, 
      author={Lajos Muzsai and David Imolai and András Lukács},
      year={2024},
      eprint={2412.01778},
      archivePrefix={arXiv},
      primaryClass={cs.CR},
      url={https://arxiv.org/abs/2412.01778}, 
}

@misc{rong2024disassemblingobfuscatedexecutablesllm,
      title={Disassembling Obfuscated Executables with LLM}, 
      author={Huanyao Rong and Yue Duan and Hang Zhang and XiaoFeng Wang and Hongbo Chen and Shengchen Duan and Shen Wang},
      year={2024},
      eprint={2407.08924},
      archivePrefix={arXiv},
      primaryClass={cs.CR},
      url={https://arxiv.org/abs/2407.08924}, 
}

@inproceedings{she2024wadecDecompilingWebassembly,
author = {She, Xinyu and Zhao, Yanjie and Wang, Haoyu},
title = {WaDec: Decompiling WebAssembly Using Large Language Model},
year = {2024},
isbn = {9798400712487},
publisher = {Association for Computing Machinery},
address = {New York, NY, USA},
url = {https://doi.org/10.1145/3691620.3695020},
doi = {10.1145/3691620.3695020},
booktitle = {Proceedings of the 39th IEEE/ACM International Conference on Automated Software Engineering},
pages = {481–492},
numpages = {12},
location = {Sacramento, CA, USA},
series = {ASE '24}
}

@inproceedings{chen2025clearagent,
  title={ClearAgent: Agentic Binary Analysis for Effective Vulnerability Detection},
  author={Chen, Xiang and Zhou, Anshunkang and Ye, Chengfeng and Zhang, Charles},
  booktitle={Proceedings of the 1st ACM SIGPLAN International Workshop on Language Models and Programming Languages},
  pages={130--137},
  year={2025}
}

@misc{GitHubRadare2,
	author = {},
	title = {{G}it{H}ub - radareorg/radare2: {U}{N}{I}{X}-like reverse engineering framework and command-line toolset --- github.com},
	howpublished = {\url{https://github.com/radareorg/radare2}},
	year = {}
}

@misc{llvmClangStatic,
	author = {},
	title = {{C}lang {S}tatic {A}nalyzer --- clang-analyzer.llvm.org},
	howpublished = {\url{https://clang-analyzer.llvm.org/}},
	year = {}
}

@inproceedings{zhou2022your,
  title={What your firmware tells you is not how you should emulate it: A specification-guided approach for firmware emulation},
  author={Zhou, Wei and Zhang, Lan and Guan, Le and Liu, Peng and Zhang, Yuqing},
  booktitle={Proceedings of the 2022 ACM SIGSAC Conference on Computer and Communications Security},
  pages={3269--3283},
  year={2022}
}

@inproceedings{qin2021idev,
  title={iDEV: Exploring and exploiting semantic deviations in ARM instruction processing},
  author={Qin, Shisong and Zhang, Chao and Chen, Kaixiang and Li, Zheming},
  booktitle={Proceedings of the 30th ACM SIGSOFT International Symposium on Software Testing and Analysis},
  pages={580--592},
  year={2021}
}

@inproceedings{west2024picture,
  title={A picture is worth 500 labels: A case study of demographic disparities in local machine learning models for Instagram and TikTok},
  author={West, Jack and Thiemt, Lea and Ahmed, Shimaa and Bartig, Maggie and Fawaz, Kassem and Banerjee, Suman},
  booktitle={2024 IEEE Symposium on Security and Privacy (SP)},
  pages={369--387},
  year={2024},
  organization={IEEE}
}

@article{cao2025jnfuzz,
  title={JNFuzz-Droid: a lightweight fuzzing and taint analysis framework for native code of Android applications},
  author={Cao, Jianchao and Guo, Fan and Qu, Yanwen},
  journal={Empirical Software Engineering},
  volume={30},
  number={5},
  pages={113},
  year={2025},
  publisher={Springer}
}

@misc{sourcewareGDBProject,
	author = {},
	title = {{G}{D}{B}: {T}he {G}{N}{U} {P}roject {D}ebugger --- sourceware.org},
	howpublished = {\url{https://www.sourceware.org/gdb/}},
	year = {}
}

@misc{abramovich2025enigmainteractivetoolssubstantially,
      title={EnIGMA: Interactive Tools Substantially Assist LM Agents in Finding Security Vulnerabilities}, 
      author={Talor Abramovich and Meet Udeshi and Minghao Shao and Kilian Lieret and Haoran Xi and Kimberly Milner and Sofija Jancheska and John Yang and Carlos E. Jimenez and Farshad Khorrami and Prashanth Krishnamurthy and Brendan Dolan-Gavitt and Muhammad Shafique and Karthik Narasimhan and Ramesh Karri and Ofir Press},
      year={2025},
      eprint={2409.16165},
      archivePrefix={arXiv},
      primaryClass={cs.AI},
      url={https://arxiv.org/abs/2409.16165}, 
}

@misc{yang2024sweagentagentcomputerinterfacesenable,
      title={SWE-agent: Agent-Computer Interfaces Enable Automated Software Engineering}, 
      author={John Yang and Carlos E. Jimenez and Alexander Wettig and Kilian Lieret and Shunyu Yao and Karthik Narasimhan and Ofir Press},
      year={2024},
      eprint={2405.15793},
      archivePrefix={arXiv},
      primaryClass={cs.SE},
      url={https://arxiv.org/abs/2405.15793}, 
}

@misc{enigmaPlusDebugsh,
	author = {},
	title = {{C}yber-{Z}ero/enigma-plus/config/commands/debug.sh at main · amazon-science/{C}yber-{Z}ero --- github.com},
	howpublished = {\url{https://github.com/amazon-science/{C}yber-{Z}ero/blob/main/enigma-plus/config/commands/debug.sh}},
	year = {}
}

@misc{triedman2025multiagentsystemsexecutearbitrary,
      title={Multi-Agent Systems Execute Arbitrary Malicious Code}, 
      author={Harold Triedman and Rishi Jha and Vitaly Shmatikov},
      year={2025},
      eprint={2503.12188},
      archivePrefix={arXiv},
      primaryClass={cs.CR},
      url={https://arxiv.org/abs/2503.12188}, 
}

@inproceedings{engel2024decidability,
  title={On the decidability of disassembling binaries},
  author={Engel, Daniel and Verbeek, Freek and Ravindran, Binoy},
  booktitle={International Symposium on Theoretical Aspects of Software Engineering},
  pages={127--145},
  year={2024},
  organization={Springer}
}

@inproceedings{dramko2025quantifying,
  title={Quantifying and Mitigating the Impact of Obfuscations on Machine-Learning-Based Decompilation Improvement},
  author={Dramko, Luke and B{\"o}l{\"o}ni-Turgut, Deniz and Le Goues, Claire and Schwartz, Edward},
  booktitle={International Conference on Detection of Intrusions and Malware, and Vulnerability Assessment},
  pages={244--266},
  year={2025},
  organization={Springer}
}

@article{triedman2025multi,
  title={Multi-agent systems execute arbitrary malicious code},
  author={Triedman, Harold and Jha, Rishi and Shmatikov, Vitaly},
  journal={arXiv preprint arXiv:2503.12188},
  year={2025}
}

@article{afianian2019malware,
  title={Malware dynamic analysis evasion techniques: A survey},
  author={Afianian, Amir and Niksefat, Salman and Sadeghiyan, Babak and Baptiste, David},
  journal={ACM Computing Surveys (CSUR)},
  volume={52},
  number={6},
  pages={1--28},
  year={2019},
  publisher={ACM New York, NY, USA}
}

@misc{frida,
	author = {},
	title = {{F}rida • {A} world-class dynamic instrumentation toolkit --- frida.re},
	howpublished = {\url{https://frida.re/}},
	year = {}
}

@misc{dynamorio,
	author = {},
	title = {dynamorio.org},
	howpublished = {\url{https://dynamorio.org/}},
	year = {},
}

@misc{androidApplicationFundamentals,
	author = {},
	title = {{A}pplication fundamentals  |  {A}pp architecture  |  {A}ndroid {D}evelopers --- developer.android.com},
	howpublished = {\url{https://developer.android.com/guide/components/fundamentals}},
	year = {},
}

@misc{wang2025salt4decompileinferringsourcelevelabstract,
      title={SALT4Decompile: Inferring Source-level Abstract Logic Tree for LLM-Based Binary Decompilation}, 
      author={Yongpan Wang and Xin Xu and Xiaojie Zhu and Xiaodong Gu and Beijun Shen},
      year={2025},
      eprint={2509.14646},
      archivePrefix={arXiv},
      primaryClass={cs.SE},
      url={https://arxiv.org/abs/2509.14646}, 
}

@misc{gem5gem5Gem5,
	author = {},
	title = { 		gem5: {T}he gem5 simulator system 	 --- gem5.org},
	howpublished = {\url{https://www.gem5.org/}},
	year = {},
}

\end{document}